\documentclass[twocolumn]{aastex631}

\usepackage{amsmath} 

\usepackage{hyperref}
\usepackage{verbatim}
\usepackage{xspace}

\newcommand\req{R_{\rm eq}}
\newcommand\ru{R_{\rm U}}
\newcommand\pu{P_{\rm U}}

\newcommand\rhoone{\rho_1}

\newcommand\rhoc{\rho_{\rm c}}
\newcommand\xice{X_{\rm ice}}
\newcommand\xcice{X_{\rm c,ice}}
\newcommand\nmoi{\rm NMoI}

\shorttitle{Uranus Gravity Science}
\shortauthors{Mankovich et al.}

\begin{document}

\title{Isolating the gravitational influence of Uranus's winds requires close passages inward of the rings}

\author{Christopher R. Mankovich}
\author{Marzia Parisi}
\author{Damon F. Landau}
\affiliation{Jet Propulsion Laboratory, California Institute of Technology}
\author{Janosz W. Dewberry}
\affiliation{Canadian Institute for Theoretical Astrophysics}
\affiliation{University of Massachussetts Amherst}

\correspondingauthor{Chris Mankovich}
\email{cmankovich@ucsc.edu}

\begin{abstract}
Close orbits by a Uranus Orbiter and Probe (UOP) could be used to deduce Uranus's multipolar gravity field to higher precision and angular degree than the $J_2$ and $J_4$ {currently measured} from ground-based ring occultations and the Voyager~2 flyby.
We examine $J_n$ sensitivity limits {obtained from} simulations of candidate UOP trajectories, pairing these with Uranus interior and wind models to perform retrievals from the gravity moments. We consider zonal wind profiles derived from recent feature-tracking data, assuming that zonal winds extend into the planet along cylinders, with a radial decay function similar to those that explain Jupiter and Saturn gravity. 
Present knowledge of $J_2$ and $J_4$ permits a fairly wide range of possible wind depths in Uranus, up to 1,800 km or $7\%$ of the planet's radius. Measuring additional gravity moments is essential to separate this unknown wind depth from other interior properties of interest, but $J_6$ is found to be {too dominated by bulk rotation to be a useful} probe of the wind depth. Odd moments arising from {Uranus's observed north-south asymmetric flow} are strong functions of the wind depth, but the usefulness of $J_3$ is hindered by its sensitivity to {present uncertainties in the wind profile}. The even moment $J_8$, or the odd moments $J_5$ and $J_7$, are the best probe{s} of the depth of Uranus's winds{. $J_8$, and most likely $J_5$ and $J_7$, are} measurable in a {highly inclined} orbit making $\gtrsim10$ pericenter passages inward of the $\zeta$ ring, approximately $1,000-2,500$ km above Uranus's cloud tops. 
\end{abstract}

\section{Introduction}\label{sec.intro}
A gravity experiment by the Uranus Orbiter and Probe (UOP) would use radio tracking of the orbiter to measure the gravitational multipole moments generated by Uranus's rotational flattening. The low angular degree ($n$) zonal moments $J_2$ and $J_4$ could be measured orders of magnitude more precisely than they have been using Voyager 2 flyby data and ground-based stellar occultations of the rings (\citealt{2024Icar..41115957F}; see also \citealt{2014AJ....148...76J,2025AJ....169...65J}). Higher-degree moments including $J_6$ and $J_8$ may be measured for the first time \cite{2024PSJ.....5..116P}. 

Similar gravity experiments by Juno and Cassini have shown that zonal wind patterns make important contributions to these moments in Jupiter and Saturn, particularly in the odd moments $n\geq3$ \citep{2018Natur.555..220I,2018Natur.555..223K,2019GeoRL..46..616G} and in the even moments with $n\gtrsim6$ \citep{2018Natur.555..227G,2019Sci...364.2965I}. These wind-induced disturbances to the planetary shape and gravity field have led to estimates for the zonal wind depth in both planets \citep{2021MNRAS.501.2352G}. \cite{2013Natur.497..344K} argued based on measurements of Uranus's $J_4$ \citep{2014AJ....148...76J} that zonal winds on Uranus are more limited in their radial extent at $d\lesssim1,000$ km, but this result is model-dependent and recent models fit to the same data allow for somewhat deeper winds \citep{2025PSJ.....6...70M}. A similar argument for shallow winds in Uranus and Neptune was made by \cite{2023AJ....165...27S}, but it is necessary to re-evaluate this question in light of more up-to-date measurements of Uranus's zonal wind profile, as well as the possibility of measuring higher degree gravity moments with a Uranus orbiter. \cite{2025PSJ.....6...27L} did not explicitly treat winds in their planetary structure models, but did conclude that proximal orbits to measure $J_6$ and possibly $J_8$ would be necessary to resolve any deep atmospheric winds, and recognized that the closest orbits proposed by \cite{2024PSJ.....5..116P} would be sufficient for this purpose. The uncertain balance between the oblateness contributions coming from rigidly rotating background structure, and the so-called dynamic contributions coming from the winds, introduces degeneracy into the retrieval of Uranus's interior structure from gravity data. It is this degeneracy that we aim to directly contend with here.

Meanwhile, studies of possible UOP trajectories have been carried out by \cite{2024PSJ.....5..116P}, who found that eccentric, nearly polar-inclination orbits with periapse passages near Uranus's equator and within the main rings are the best means of constraining the even zonal gravity moments $J_2$--$J_8$, assuming at least 8 total gravity passes. Their alternative trajectory outside the rings was found to still dramatically improve $J_2$ and $J_4$ relative to \cite{2014AJ....148...76J}, but would be unlikely to detect $J_6$ or $J_8$ above the noise floor, a result of the steeper power-law decay of higher $n$ components of the potential as a function of distance from Uranus. (The stronger north-south asymmetry of this type of orbit also exacerbates the poor sensitivity to the equatorially concentrated zonal moments.) Even arbitrarily good precision on $J_2$ and $J_4$ cannot alleviate the interior/wind degeneracy described above.

With the goal of informing trajectory planning for UOP, here we investigate a broader range of possible interior structure and wind configurations for Uranus, aiming to better understand these confounding variables and the precision required of these gravity moments. Even Uranus's cloud-top circulations are not fully understood, with different data sets indicating tropospheric flow that is a nontrivial function of latitude, and appears to exhibit some north-south asymmetry \citep{2015Icar..258..192S,2015Icar..250..294K,2024Icar..42016186S}. We address the question of whether this asymmetry may be detectable by UOP via measurements of the odd zonal moments $J_3$, $J_5$, etc. We study how the possible wind contributions to the even and odd $J_n$ affect our ability to deduce interior quantities of interest, such as the density or composition of Uranus's core. Section 2 of this paper describes the methods used to model Uranus's interior, zonal velocity profile, and gravity field. Section 3 compares our modeled gravity moments to sensitivity limits for present (Voyager- and ground-based) and simulated (orbiter-based) gravity field measurements. Section 4 presents simulated retrievals to improved measurements of the gravity moments that could be made by an orbiter. Section 5 presents a discussion and Section 6 summarizes our findings.

\section{Uranus's gravity field}\label{sec.methods}
We model the Uranian gravity field using the same approach as in \cite{2025PSJ.....6...70M} (hereafter M25), first computing oblate reference models assuming rigid rotation and using the theory of figures \citep{2017A&A...606A.139N,2021PSJ.....2..241N}, then adopting {an axially} symmetric wind profile and calculating the modifications to the shape and gravitational potential by solving for thermo-gravitational wind balance (\citealt{2023PSJ.....4...59M}; see also \citealt{2017JFM...810..175G} and \citealt{2017JGRE..122..686C}). Unlike M25 where wind-induced perturbations to the $J_n$ were pre-computed and interpolated during production runs, here all wind solves are carried out on the fly{, eliminating the small $J_n$ residuals introduced by interpolation using precomputed models (see M25 Figure~18)}.

\subsection{Rigid rotation and oblateness}\label{sec.methods.interior}

To model Uranus's interior we adopt the same polytrope-based structure as in M25, here considering only `gradient' models that incorporate a smooth transition from an envelope to a less compressible core. An atmosphere break is introduced at $r=0.9\ru$ with the atmospheric polytropic index fixed at $n_{\rm poly}=2$ such that realistic densities are recovered near $P=1$ bar (see discussion in M25). These models are agnostic as to the phase, composition, and temperature of the material in Uranus's interior, but post-hoc comparisons (e.g., to the approximate density of known mixtures) will aid in contextualizing our results. {While the models chosen here provide a broad range of density profiles, they are by no means intended to exhaust the full space of interior structures permissible for Uranus, which may include one or more density interfaces or multiple composition gradients. The `gradient' and `interface' models from M25 yield  very similar posterior distributions for Uranus's wind depth, and hence either class of model is suitable for our primary focus here: to study which $J_n$ are needed to disentangle the gravitational influence of Uranus's winds from that of its internal structure.}

Our baseline models are constrained only by $J_2$, $J_4$, and $\rhoone$, with six free parameters sampled using \verb+emcee+\footnote{\url{https://emcee.readthedocs.io/en/v3.1.6/}} \citep{2013PSP..125..306F}. These parameters specify the inner and outer radii of the gradient region $r_i$ and $r_o$, the core density enhancement factor $\alpha$, the bulk spin period $\pu$, the rigid body degree-2 zonal gravity moment $J_2^{\rm rigid}$, and the wind depth $d$. (The wind model is described in the following section.) Uranus's mass $8.68129\times10^{28}$ g and equatorial radius $\req=25,559$ km are satisfied by construction. The total $J_2$ and $J_4$ are fit directly to the \cite{2024Icar..41115957F} values. For simplicity we neglect the strong correlation between $J_2$ and $J_4$ when calculating the model likelihood. This likely leads us to overestimate the variance of the posterior distributions somewhat; see M25 for similar retrievals that account for this correlation but still manifest all the major degeneracies relevant to the present discussion. 

The oblateness and zonal gravity moments associated with the reference rigid-body structure 
are computed from the fourth-order theory of figures (ToF4){, an approximation that yields systematically biased $J_{2n}$ compared to more accurate methods (e.g., \citep{2018Natur.555..227G})}. To ensure the reliability of this method, we compare to the more accurate, much slower seventh-order algorithm (ToF7; \citealt{2021PSJ.....2..241N}) for 1,000 models {randomly drawn from our baseline sample}. We find errors (ToF7 minus ToF4) are narrowly distributed around $-5\times10^{-11}$ for $J_2$, $-3\times10^{-7}$ for $J_4$, $+9\times10^{-9}$ for $J_6$, and $+10^{-10}$ for $J_8$. 
{Among these, the $J_4$ error is of possible concern, only $\sim50\%$ smaller in magnitude than the \cite{2016Icar..279...62F} $1\sigma$ uncertainty we sample from. To quantify any bias arising from this $J_4$ error, we obtained a sample where the \cite{2016Icar..279...62F} $J_4$ centroid was artificially increased by $+3\times10^{-7}$, capturing the main effect of a switch to the more accurate ToF7 while circumventing its prohibitive computational expense. The full model and sampling process is described below, but here it suffices to say that based on this test, the ToF4 approximation only subtly alters the posterior solution space overall. The largest effect is that the $J_4$ error associated with ToF4 favors marginally smaller wind depths, underestimating the median wind depth by $\sim70$ km compared to ToF7. This bias is small compared to the $\sim1,500$ km range of possible wind depths and we verified that it does not affect the detectability of the $J_n$ at the level of precision reported in Section~\ref{sec.results} below.} 
At higher degree, we find $J_8$ errors one to two orders of magnitude smaller than the $3\sigma$ uncertanties on $J_8$ in the most optimistic UOP trajectories considered below. In all cases, these errors introduced by the fourth-order approximation are negligible compared to the total value of each model's gravity moment, and therefore the faster ToF4 is sufficiently accurate for our purposes.

\subsection{Wind model}\label{sec.methods.winds}
We make the assumption that the wind velocities at Uranus's cloud tops can be related to velocities in the deeper atmosphere by first extending the velocity along cylinders coaxial with Uranus's spin axis, then applying a decay of the differential rotation as a function of radial distance from Uranus's oblate 1-bar surface, yielding a differential flow velocity
\begin{equation}\label{eq.wind_speed_interior}
    u(r,\mu)=u_{\rm surf}(\tilde\mu)f(r,\mu).
\end{equation} 
Here $\mu=\cos\theta$, $u_{\rm surf}$ is the azimuthal velocity derived from feature tracking (with rigid body rotation subtracted), $\tilde\mu=\sqrt{1-[r/r_{\rm surf}(\mu)]^2(1-\mu^2)}$ is the cosine latitude of a surface point sharing a common cylindrical radius with the interior point $(r,\mu)$, and the decay function $f$ is taken to be a sigmoid function
\begin{equation}\label{eq.wind_decay_function}
    f(r,\mu) = \frac12\left(1+\tanh\left[\frac{r-[r_{\rm surf}(\mu)-d]}{w}\right]\right)
\end{equation}
following \cite{2021PSJ.....2..198D} and motivated by the radially thin shear zones found for Jupiter and Saturn by \cite{2021MNRAS.501.2352G}.
Here $r_{\rm surf}(\mu)$ is the oblate shape of the model outer boundary, {the decay width} $w$ is taken to be constant at 0.02 to yield a thin shear zone, and the decay centroid depth $d$ is a free parameter. The assumption that the wind penetration depth is independent of latitude may prove ill-founded when better gravity field data become available, as \cite{2025NatCo..16.2618G} have shown for Saturn based on Cassini data. Note that because the winds alter the hydrostatic balance, they in fact alter $r_{\rm surf}(\mu)$ as well, but this minor effect is neglected here, $r_{\rm surf}(\mu)$ being taken from the rigidly rotating theory of figures model.

We choose a uniform prior for wind depths $d$ between $10^{-2}~\req=256$~km and $10^{-1}~\req=2560$ km. All other priors are as described in M25. We find the nonzero lower limit for $d$ to be necessary to avoid inefficient MCMC sampling close to $d=0$ where the wind perturbations to $J_2$ and $J_4$ vanish, causing the likelihood to become insensitive to $d$. Additionally, a reliable treatment of very shallow winds would require higher spatial resolution in our wind solve than is practical for the $\mathcal O(10^5)$ likelihood evaluations carried out here for each retrieval. {Here TGWE solves use 40 quasi-radial zones (see \citealt{2023PSJ.....4...59M} Appendix~A for details), which resolution tests show is sufficient to resolve all the $J_{n}^{\rm winds}$ to an absolute error less than $10^{-8}$ for the present case. The moments $J_2$ and $J_4$ are the most resolution-dependent by a wide margin; at our adopted resolution, the $n=3$ and $n\geq5$ moments have absolute errors less than $10^{-10}$.}

\begin{deluxetable*}{lll}
    \tablecaption{Adopted cloud-level wind profiles \label{tab.wind_fits}}
    \tabletypesize{\scriptsize}
    \tablewidth{0pt} 
    \tablehead{
        \colhead{Label} & \colhead{Description}& \colhead{Source$^\dag$}
        }
    \startdata
    Even & North-south symmetric & \S4.2; Equations (2-4) and Table 5 \\
    Asym. 5 & Mid-latitude asymmetry & \S4.3; Eq. (5) \\
    Asym. 6 & Mid-latitude asymmetry excluding 2012-2014 & \S4.3; Eq. (6) \\
    Composite & Pole-to-pole composite including Voyager southern latitudes  & \S4.5 and Tables 6-7 \\
    \enddata
\tablecomments{
    $^\dag$Sources here refer to \cite{2015Icar..258..192S}. 
    }
\end{deluxetable*}

\begin{figure}
    \begin{center}
        \includegraphics[width=\columnwidth]{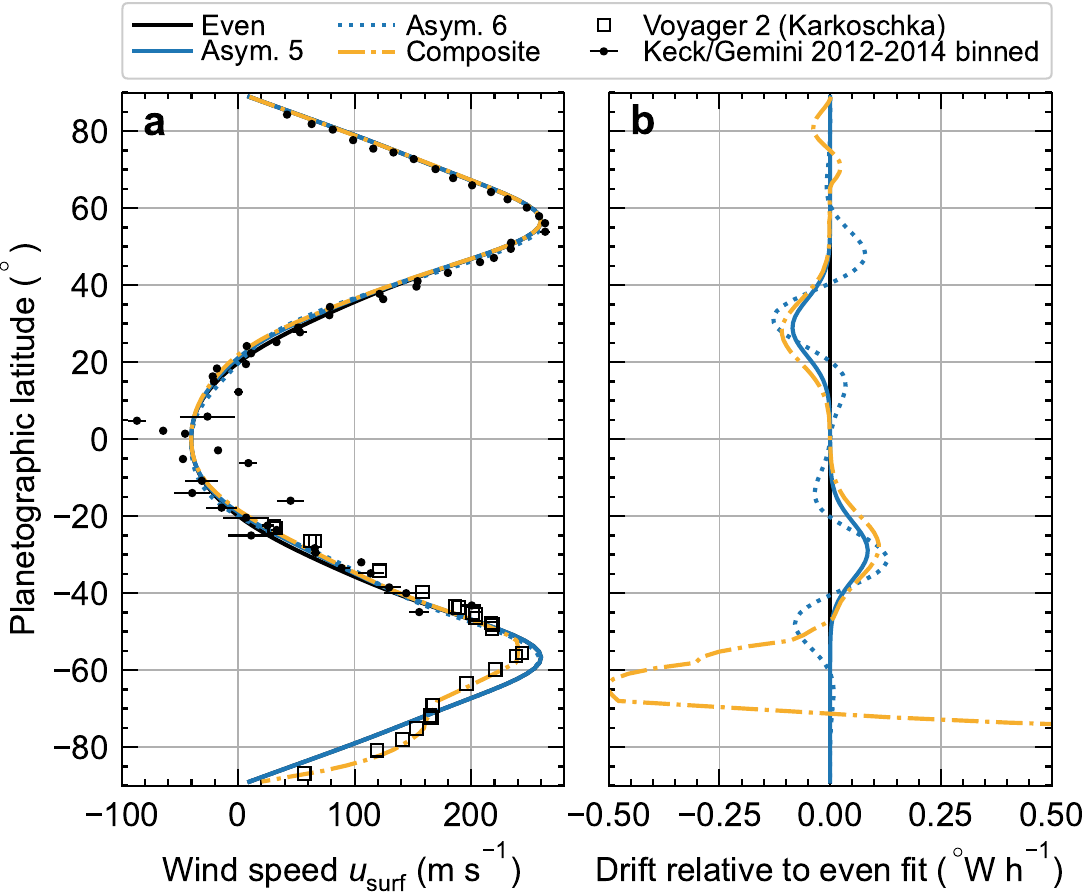}
        \caption{\label{fig.baseline_best_windprofs}
        {Different data sets tracking Uranus's tropospheric features yield a range of permitted zonal wind profiles.}
         Panel (a) gives wind speed as a function of planetographic latitude for the even fit (black) and three asymmetric fits (colors). Panel (b) {examines north-south asymmetry, giving} the angular {velocity difference} ($^\circ$ west per hour) relative to the even fit for each of the three odd fits. Data and fits are from \cite{2015Icar..258..192S} and \cite{2015Icar..250..294K}; see Table~\ref{tab.wind_fits}.
        }
    \end{center}
\end{figure}

Under these assumptions the differential velocity (relative to the bulk rotation with period $\pu$) is completely specified by the free parameter $d$ and the cloud-level wind profile $u_{\rm surf}(\mu)$.
How well constrained then are the cloud-level winds as a function of latitude? \cite{2015Icar..258..192S} present a synthesis of decades of Uranus imaging, including data from Voyager, Hubble, Gemini and Keck, with coverage gradually growing from the southern hemisphere into the northern hemisphere while the data from Earth-based observations have improved in quality. Sromovsky et al. present a number of wind fits, summarized here in Table~\ref{tab.wind_fits} and Figure~\ref{fig.baseline_best_windprofs}, which plots the latitudinal profiles of wind speed, as well as the drift rate of the asymmetric profiles relative to the symmetric profile. Profiles \textit{Asym. 5} and \textit{Asym. 6} deviate from the symmetric fit in a way that is purely antisymmetric by construction, and hence imply no differences in the even $J_n$. The more strongly asymmetric \textit{Composite} fit does introduce some equatorially symmetric drift at high latitudes that can alter $J_{10}$, and to a much lesser degree $J_8$, as Section~\ref{sec.results} will show.

Because the deep atmospheric differential rotation is assumed to propagate downward from the surface along surfaces of constant distance from the rotation axis, any north-south antisymmetric components of the winds (see Figure~\ref{fig.baseline_best_windprofs}b) can lead to shear at the equator if the winds are sufficiently deep \citep{2021PSJ.....2..198D}. {We find that for the deepest winds compatible with the gravity field, an equatorial shear of magnitude up to $\sim8\ {\rm m\ s}^{-1}$ is possible, but even for these extreme depths the low-degree odd $J_n$ are affected by $\lesssim10\%$.}

It is clear from Figure~\ref{fig.baseline_best_windprofs}a that the data could also be compatible with modestly different zonal profiles not considered here, especially near the equator where the dearth of compact cloud features means wind speeds are more poorly constrained \citep{2015Icar..258..192S}. Nonetheless these four profiles serve as an appropriate basis for studying the gravitational influence of Uranus's winds given the information at hand.

\begin{deluxetable*}{cllllll}
    \tablecaption{Reference trajectories \label{tab.orbits}}
    \tabletypesize{\scriptsize}
    \tablewidth{0pt} 
    \tablehead{
        \colhead{Case} & 
        \colhead{Periapsis altitude} & 
        \colhead{Periapsis latitude} & 
        \colhead{Inclination} & 
        \colhead{Sun-Uranus-periapsis} & 
        \colhead{Ring crossing} & 
        \colhead{Notes}
        }
    \startdata
    1 & 2500 km & 45 deg & 45 deg & 100 deg & Outside $\epsilon$ & 27000 km altitude at ring crossing \\
    2 & 2500 & 0 & 30 & 80 & Inside $\zeta$ & Threshold from \cite{11068400} \\
    3 & 2500 & 0 & 80 & 20 & Inside $\zeta$ & Baseline from \cite{11068400} \\
    4 & 27000 & 0 & 80 & 20 & Outside $\epsilon$ & High periapsis, edge-on \\ 
    5 & 1000 & 0 & 60 & -22 & Inside $\zeta$ & Edge-on, mid-inclination 
    \enddata
\tablecomments{
    All cases have 10 orbits (11 periapses) with approximately $42$-day period. First periapsis is February 2054 with solar conjunction centered at apoapsis of the middle orbit.
    }
\end{deluxetable*}

\section{Results}\label{sec.results}

\begin{figure}
    \begin{center}
        \includegraphics[width=0.9\columnwidth]{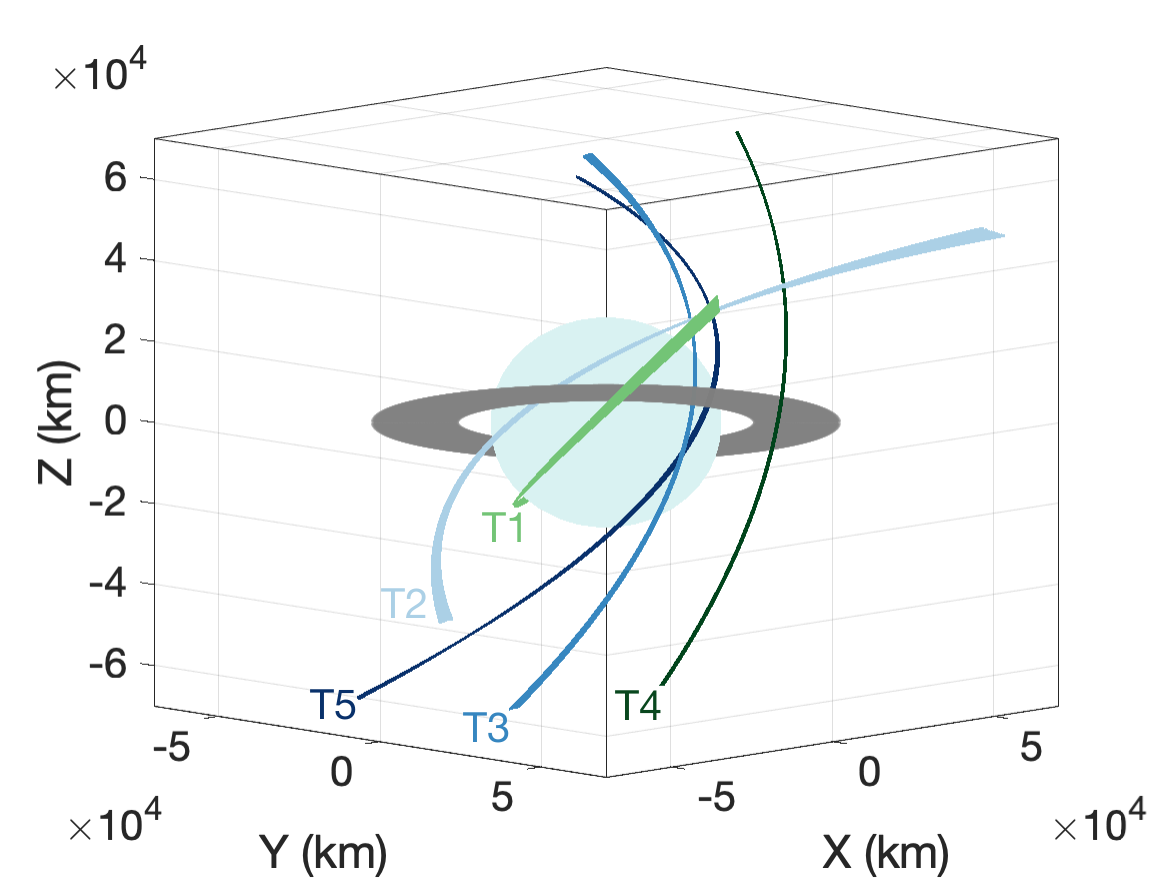}
        \caption{\label{fig.orbits}
        {
            Close passages of Uranus for each of the five trajectories listed in Table~\ref{tab.orbits}.
            The rings are shown extending from 7,000 km to 25,590 km altitude above the 1-bar surface, spanning the innermost extent of the $\zeta$ ring estimated by \cite{2006Icar..180..186D} (but see discussion in Section~\ref{sec.discussion}) through the $\epsilon$ ring.
            The viewing inclination is 10$^\circ$ below the ring plane.
        }
        }
    \end{center}
\end{figure}

We present our models in the context of uncertainties on Uranus's zonal gravity moments $J_n$ expected for a number of candidate UOP trajectories. {These trajectories are summarized in Table~\ref{tab.orbits} and Figure~\ref{fig.orbits}}. 
All have high eccentricities to enable a moon tour in addition to close passages to Uranus. Trajectories 1, 2 and 3 are similar to those presented in \cite{2024PSJ.....5..116P} but with some differences. In contrast to \cite{2024PSJ.....5..116P}, all three trajectories have approximately the same pericenter altitudes, in the range $2,600$ to $2,800$ km above Uranus's cloud tops. T1 crosses Uranus's equatorial plane outside the main rings, while T2 and T3 cross within the 4, 5, and 6 rings, a choice that risks flying closer to tenuous ring material \citep{2006Icar..180..186D} in exchange for pericenter passages closer to the planet's equator ($0^\circ$ latitude compared to T1's $-45^\circ$). Among T2 and T3, T3's higher inclination ($80^\circ$ compared to T2's $30^\circ$) lends it superior latitudinal coverage, rendering T3 the most sensitive of the three for measuring the zonal gravity moments. 

We also present two previously unreported trajectories {T4} and {T5} for comparison. {T4} is another highly inclined orbit ($80^\circ$) with closest approach {just outside the $\epsilon$ ring} at a pericenter altitude of $27,000$~km, yielding $J_n$ uncertainties slightly smaller than those from T2, but orders of magnitude larger than those expected from T3. Finally {T5} is a moderately inclined ($60^\circ$) orbit that makes very close pericenter passages ($1,000$~km altitude) above the equator, inside the rings, yielding $J_n$ sensitivity that rivals that of T3. In what follows we give quantitative comparisons of all trajectories' $J_n$ sensitivities to the moments predicted by our planetary structure models.

{It should be kept in mind that the $J_n$ uncertainties are a function of the noise level of the radio Doppler signal used to track the spacecraft, which in turn depends on the unknown amplitude of Uranus's normal modes \citep{2024PSJ.....5..116P}. Jupiter and Saturn's gravity fields both show indirect evidence for time variability, possibly from normal mode oscillations \citep{2020PSJ.....1...27M,2022NatCo..13.4632D}, and this noise source had to be taken into account in the analysis of Juno \citep{2020GeoRL..4786572D} and Cassini \citep{2019Sci...364.2965I} radio science. The uncertainties we compare to here (\cite{2024PSJ.....5..116P}'s Saturn-like) are based on the Cassini noise level that \cite{2019Sci...364.2965I} used for Cassini gravity science.}

\subsection{Will the asymmetry in the gravity field constrain the wind depth?}
\label{sec.results.odds}
\begin{figure}
    \begin{center}
        \includegraphics[width=\columnwidth]{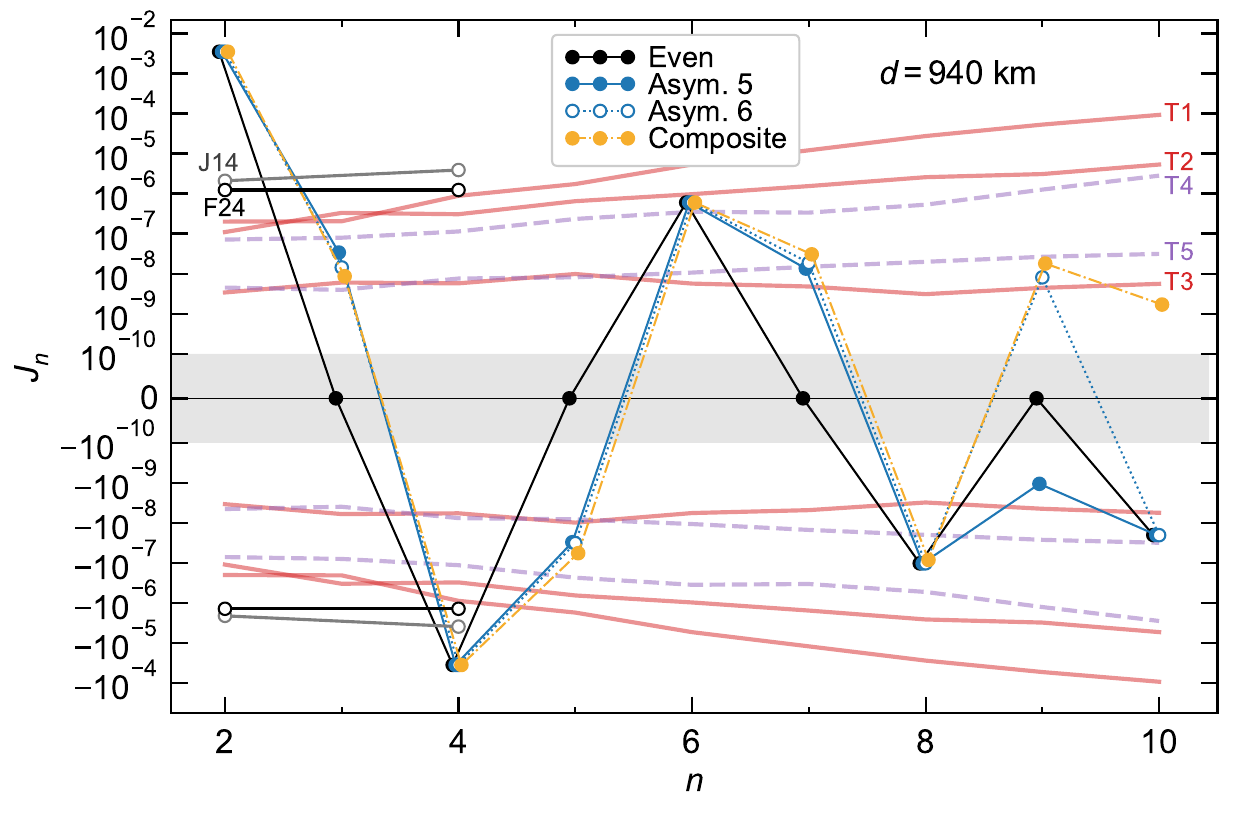}
        \caption{\label{fig.baseline_best_jn}
        Gravity moments calculated by applying each of the four wind models displayed in Figure~\ref{fig.baseline_best_windprofs} to the same interior model. The same wind radial decay function is enforced in all cases, with winds extending to a depth $d=940$ km{, close to the median value obtained assuming the Even wind fit}. The thick red lines denote the $\pm3\sigma$ detection limits for Trajectories 1, 2, and 3{; dashed purple shows those of Trajectories 4 and 5} {(see Table~\ref{tab.orbits})}. The {grey and black lines} at $n=2,4$ labeled J14 {and F24} give the \cite{2014AJ....148...76J} {and \cite{2024Icar..41115957F}} $3\sigma$ uncertainties on the measured moments $J_2$ and $J_4$.
        }
    \end{center}
\end{figure}

Figure~\ref{fig.baseline_best_jn} shows the gravity spectra that follow from applying the four wind profiles of Figure~\ref{fig.baseline_best_windprofs} and Table~\ref{tab.wind_fits} to a single interior model and decay depth $d\approx1,200$ km. 
{The moments broadly agree from one wind profile to the next. Deep asymmetric winds generate $J_3$, $J_5$, and $J_7$ all rising above their $3\sigma$ detectability limits for T3; none would yield a firm detection by T1 or T2. Consistency between these wind profiles breaks down at $J_9$, where the less pronounced mid-latitude asymmetry of the \textit{Asym. 5} fit produces a different sign compared to the other odd fits. The even harmonics behave similarly, showing good consistency until $J_{10}$, whose overall sign is dependent} on whether one adopts the \textit{Composite} fit incorporating \cite{2015Icar..250..294K}'s reanalysis of Voyager imaging of southern latitudes. Even the even component of this profile has enhanced power closer to the poles (see the yellow curve in Figure~\ref{fig.baseline_best_windprofs}b), which most strongly affects the highest angular degrees. As noted by Sromovsky et al., the data combined for this \textit{Composite} fit were obtained decades apart and it remains unclear to what degree this profile represents a persistent global state in Uranus. 

{In light of the wide variance of $J_9$ and $J_{10}$ as a function of the uncertain wind profile, these moments are not directly predictive of the wind depth. However, based on the evidence examined so far, $J_3$ through $J_8$ could be. At this $\approx1,200$ km wind depth, all have similar order of magnitude, and all are detectable in T3 but not T1 or T2. T4 performs similarly to T2, and T5 similar to T3. Hence, the orbits with low-latitude periapses inward of the rings appear to be necessary if any of the moments $J_3-J_8$ are required. In what follows, we explore a wide range of possibly wind depths and argue that these moments are essential for improving our knowledge of interior structure.}

\subsection{Retrievals from even $J_n$}\label{sec.results.retrievals}

\begin{figure*}
    \begin{center}
        \includegraphics[width=\textwidth]{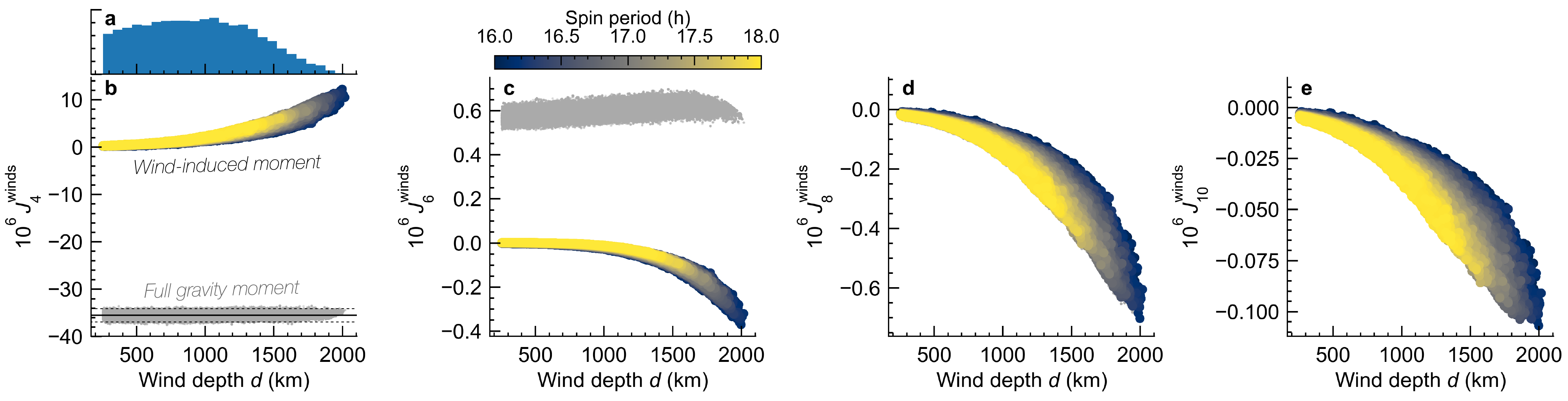}
        \caption{\label{fig.dj2n-depth-spin}
        The wind-induced even zonal moments $J_{2n}$ as a function of wind depth and bulk rotation. Rotation period is mapped to colors running from 16~h (dark blue) to 18~h (bright yellow). Panel (\textbf{a}) shows the posterior   histogram of wind depth. Panels (\textbf{b}-\textbf{e}) plot the even gravity moments from $n=2$ through $n=10$. For comparison, grey points show the distribution of full gravity moments (rigid body plus winds). The rigid-body components are negligible for $J_8$ and $J_{10}$ such that grey points fall behind the colorful points in panels \textbf{d}-\textbf{e}. The range in panel (\textbf{b}) shows the $\pm3\sigma$ range of the observational $J_4$ constraint.
        }
    \end{center}
\end{figure*}

We look beyond a best-fit model and compute a distribution of models sampled from a Gaussian likelihood in $J_2$, $J_4$, and $\rhoone$. {In what follows, we default to the fit \textit{Asym. 5} that retains the mean mid-latitude asymmetry present in observations spanning Voyager through 2014, especially in imaging from Hubble (e.g., \citealt{1998Sci...280..570K,2001Icar..153..229H}) and Keck (\citealt{2009Icar..203..265S,2009Icar..201..257H}); see \cite{2015Icar..258..192S} for further references. We repeat the sampling process for the other wind profiles for comparison, but numbers quoted in the text assume the \textit{Asym. 5} fit unless otherwise noted. We note that all wind profiles yield nearly identical distributions for the wind depth $d$.}

The data permit a broad distribution of wind depths, up to $0.072~\req\approx1,830$ km (99\% quantile). The median wind depth is $0.037~\req\approx940$ km, close to that of the maximum-likelihood model seen in Figure~\ref{fig.baseline_best_jn}. Figure~\ref{fig.dj2n-depth-spin} shows the distribution of retrieved wind depths and the corresponding wind-induced even zonal gravity moments. As in M25 these wind contributions are only weakly sensitive to the bulk rotation ($\pu$; color map in the figure) and to other parameters describing the interior (visible in the overall dispersion of points at constant $d$ and $\pu$, most obvious in panels d and e). The wind contributions are dictated mostly by the depth $d$ itself. Inspecting the total value of $J_4$ compared to the imposed observational uncertainty (Figure~\ref{fig.dj2n-depth-spin}) shows that it is the $J_4$ constraint that sets the upper limit on retrieved wind depth.

It bears emphasizing that, with $J_2$ and $J_4$ constrained by their observed values, the total value of $J_6$ is largely insensitive to the wind depth. For the grey points in Figure~\ref{fig.dj2n-depth-spin}, $J_6$ exhibits a weak correlation with $d$ (panel c; Pearson correlation coefficient $0.63$) compared to the $J_8-d$ distribution (panel d; $0.91$). {The large dispersion in $J_6$ at fixed depth, comparable to its full range over all depths, reflects this moment's sensitivity to uncertain interior structure. This sensitivity is valuable for the long-term goal of quantifying Uranus's interior structure, but it means that we} do not expect a measurement of $J_6$ alone to provide a {useful} discriminant of the depth of Uranus's zonal flow.
{(Note that the rigid-body components of $J_8$ and $J_{10}$ are negligible and hence the grey points fall behind the colorful points in Figure~\ref{fig.dj2n-depth-spin} panels d-e.)}

\begin{figure}
    \begin{center}
        \includegraphics[width=\columnwidth]{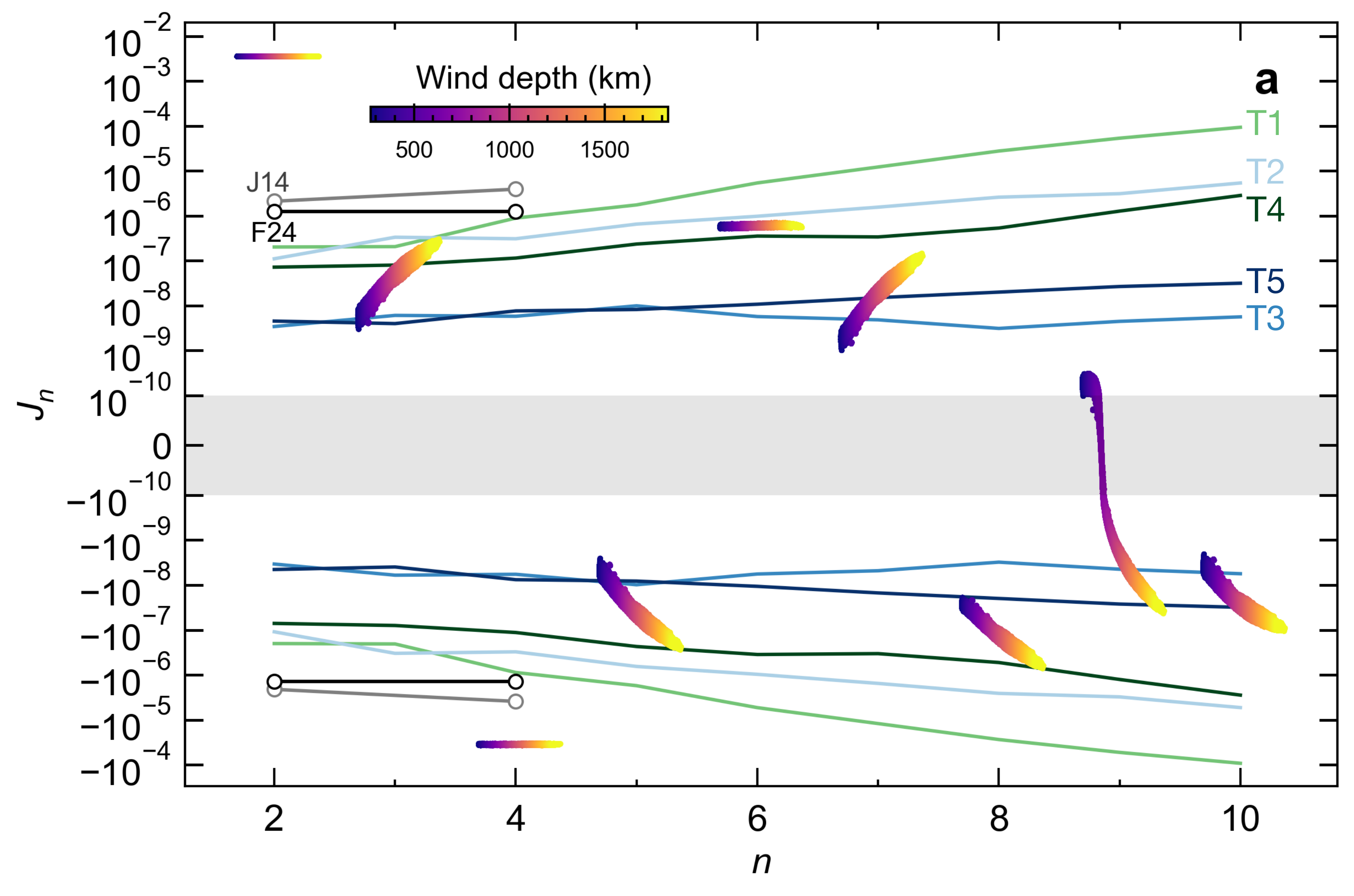} \\
        \includegraphics[width=\columnwidth]{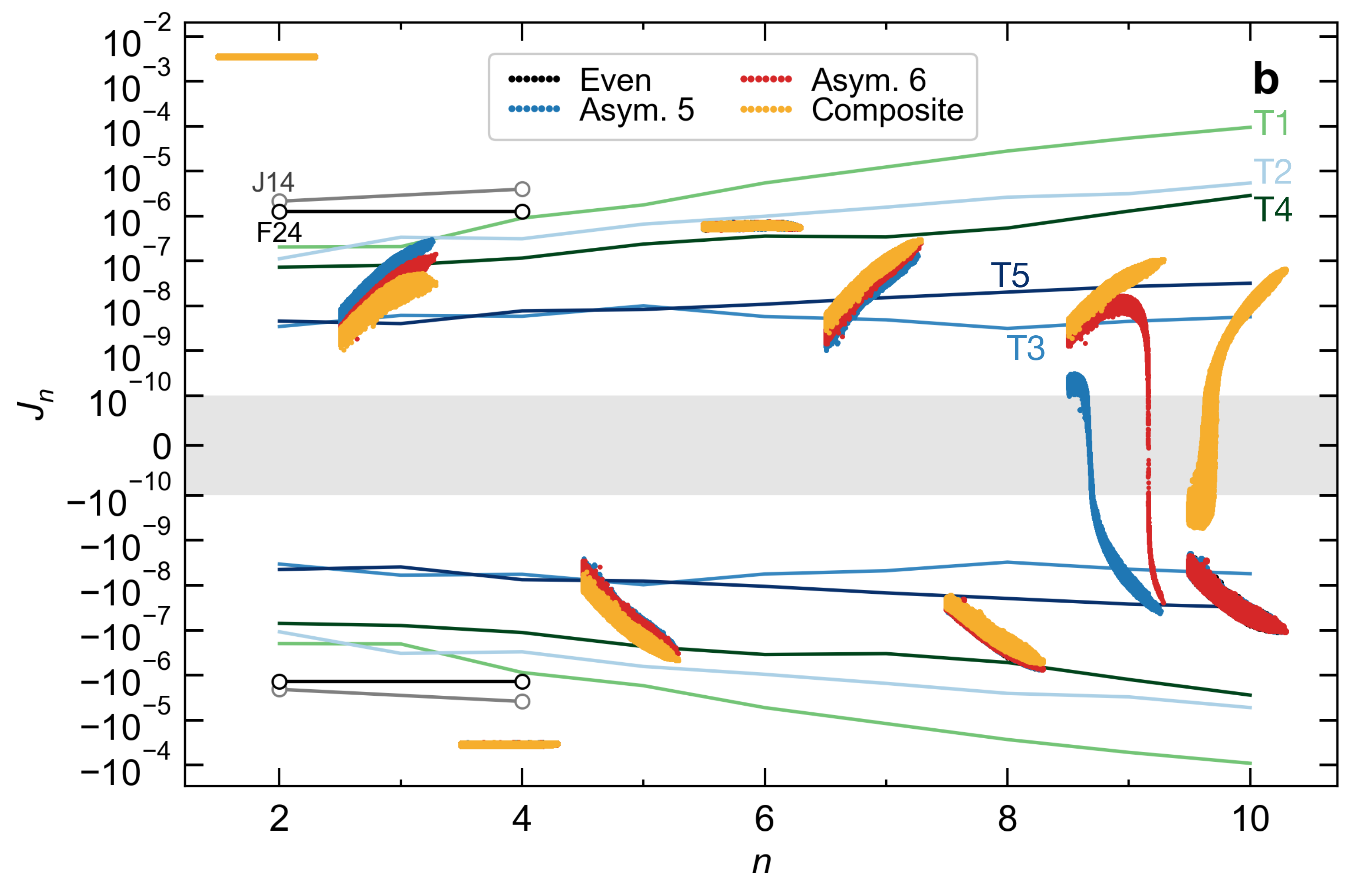}
        \caption{\label{fig.jn_all_fit_j2j4}
        Panel (\textbf{a}) shows gravity moments for the full sample of models constrained by $J_2$ and $J_4$, assuming wind profile Asym. 5, with color mapped to the spin depth $d$ (see inset colorbar). For each value of $n$, the points are also slightly spread out horizontally so that wind depth $d$ increases linearly from left to right within a single cluster. Odds are plotted with smaller symbols. The colored curves mark $3\sigma$ sensitivity for candidate UOP trajectories. {J14 and F24 mark the $3\sigma$ uncertainties on the measured moments $J_2$ and $J_4$ \citep{2014AJ....148...76J,2024Icar..41115957F}.} The y axis switches from logarithmic to linear scale in the range $\pm10^{-10}$ (grey shaded region).
        Panel (\textbf{b}) compares different samples obtained assuming each of the four wind profiles in turn; see Table~\ref{tab.wind_fits}. Here for each $n$ value, the points are spread horizontally such that depth increases linearly from $0.01~\req=256$ km at $n-\frac12$ to $0.10~\req=2560$ km at $n+\frac12$.
        {Black points are omitted at odd $n$ where even winds generate no moment. At even $n$, black and blue points are obscured by red points, since the differences between these fits are purely odd and hence do not affect the even $J_n$.}
        }
    \end{center}
\end{figure}

\begin{figure}
    \begin{center}
        \includegraphics[width=\columnwidth]{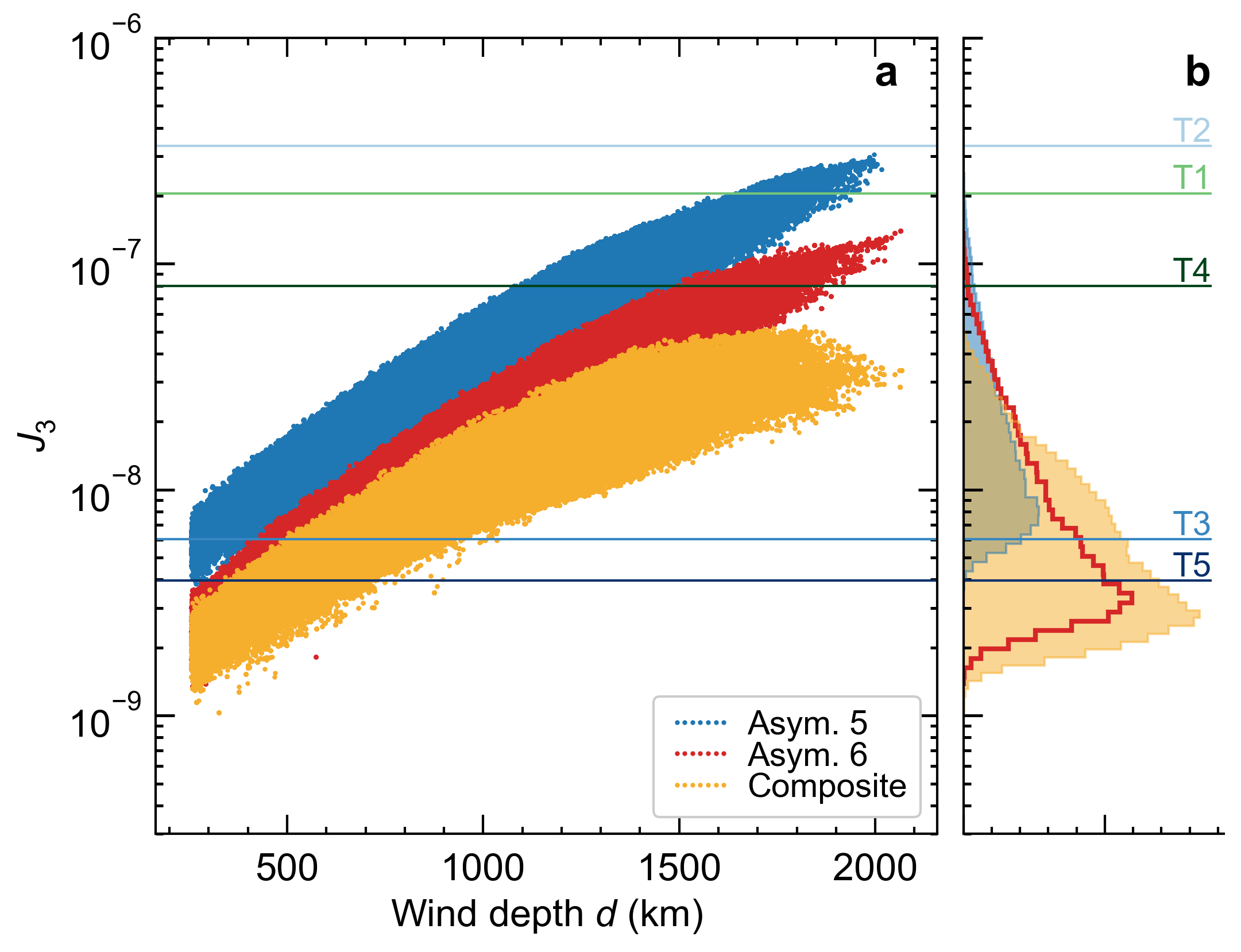}
        \caption{\label{fig.j3}
        {$J_3$ as a function of wind depth (\textbf{a}) and $J_3$ histograms (\textbf{b}) for three wind profiles with different degrees of equatorial asymmetry. Horizontal lines mark the $3\sigma_{J_3}$ detection limits for the five trajectories in Table~\ref{tab.orbits}; in each scenario, points falling below the line would evade detection. The assumed wind profile bears strongly on the value of $J_3$, making it difficult to uniquely constrain the wind depth using $J_3$ alone.}
        }
    \end{center}
\end{figure}

Figure~\ref{fig.jn_all_fit_j2j4}a summarizes the spectrum of full gravity moments through $J_{10}$ for the same sample. 
The odd zonal moments are strong functions of the wind depth{, as expected.}
Among the even zonal moments, only at $n=8$ does the differential rotation contribution become comparable in magnitude to the rigid-body contribution, giving rise to a strong depth dependence in the total gravity moment. The magnitude of $J_8$ meets or exceeds the $3\sigma$ threshold for detectability in Trajectories 3 and 5 for virtually all wind depths, even the most modest cases at $d=0.01~\req=256$ km. 
In fact, for the deepest winds $d\gtrsim0.066~\req\approx1,700$~km, $J_8$ does rise above the detection threshold even in T4, whose {closest approach lies more conservatively outside the $\epsilon$ ring}. 
However, these depths are rather extreme, only attained by 1\% of the sample. We therefore underscore that pericenter passages within the rings, where the orbiter can pass closer to Uranus's cloud tops (as in T3 or T5), remain the surest means of gaining information about the depth of Uranus's winds by making a definitive measurement of $J_8$. 
{$J_{10}$ is a more fraught proposition until Uranus's tropospheric flow is better understood, for the reasons discussed in Section~\ref{sec.results.odds}.}

Figure~\ref{fig.jn_all_fit_j2j4}{b} similarly plots the gravity spectra $J_n$ but compares the four wind profiles directly, generalizing Figure~\ref{fig.baseline_best_jn} to show all wind depths permitted by present knowledge of Uranus's gravity field. The values of $J_6$ and $J_8$ appear to be robust with respect to the wind profile assumed, as expected given that the profiles differ mostly in their degree of asymmetry about the equator. Among these moments, $J_8$ exhibits the strongest dependence on wind depth, evident in the slope of the $n=8$ cluster in the diagram. The variance in wind depth yields an overall spread in $J_8$ values that exceeds the variance that follows from the uncertain interior parameters (the spread in $J_8$ at fixed depth). Meanwhile $J_{10}$ is a similarly strong function of depth, but is weaker by approximately an order of magnitude or more, and its value and even sign varies depending on whether the Voyager-based measurements of deep southern wind speeds are included (gold versus the other $J_{10}$ points). These results confirm our finding that $J_8$ is the even moment most likely to reveal Uranus's zonal wind depth. 

{Turning our attention to the odd $J_n$, the more pronounced slope of the $J_3$, $J_5$, and $J_7$ clusters reveal these moments' strong sensitivity to the wind depth. Leaving aside $J_9$ because of its overwhelming dependence on the assumed wind profile, we find that even $J_3$ is perhaps uncomfortably affected by these same uncertainties: Figure~\ref{fig.j3} shows that $J_3$ can vary by as much as an order of magnitude at fixed wind depth, depending on which wind profile is assumed. A comparison of the $J_3$ histograms to the T1-T5 sensitivity limits suggests that $J_3$ may not be detectable from even the more optimistic orbits. Assuming T3 flies, the wind fit \textit{Asym. 5} produces detectable $J_3$ in 98\% of models, dropping to 71\% if the \textit{Composite} fit is adopted instead. If the more conservative T1 flies, only 2\% of models would generate detectable $J_3$ assuming the \textit{Asym 5.} profile, and this fraction drops to zero for the other profiles. We conclude that $J_3$ on its own does not uniquely constrain the wind depth, at least until Uranus's global circulation is better characterized.}

{Generally speaking, the shallowest winds are the most likely to evade detection, and hence a null result for one or more of the odd $J_n$ could offer a useful upper limit for the truncation depth of the winds. This is only possible provided an orbit like T3 or T5 with a low enough noise level that a null detection would rule out a majority of the models seen in Figures~\ref{fig.jn_all_fit_j2j4}b or~\ref{fig.j3}. But for these same orbits, a measurement of $J_8$ is expected and its magnitude would offer the best constraint on the extent of the zonal flow. For the more conservative T1, T2, or T4, a null detection of the odd moments is the most likely outcome and little to no information would be gained.}
Note that taken at face value, Figure~\ref{fig.jn_all_fit_j2j4} suggests the odd $J_n$ are necessarily different from zero, but this is merely an artifact of the nonzero lower limit imposed on depth $d$ for numerical reasons described in Section~\ref{sec.methods.interior}. Physically, as $d$ approaches zero, the odd moments vanish and the evens approach an approximately power-law distribution dictated by the rigid-body rotation.

\begin{figure}[ht!]
    \begin{center}
        \includegraphics[width=\columnwidth]{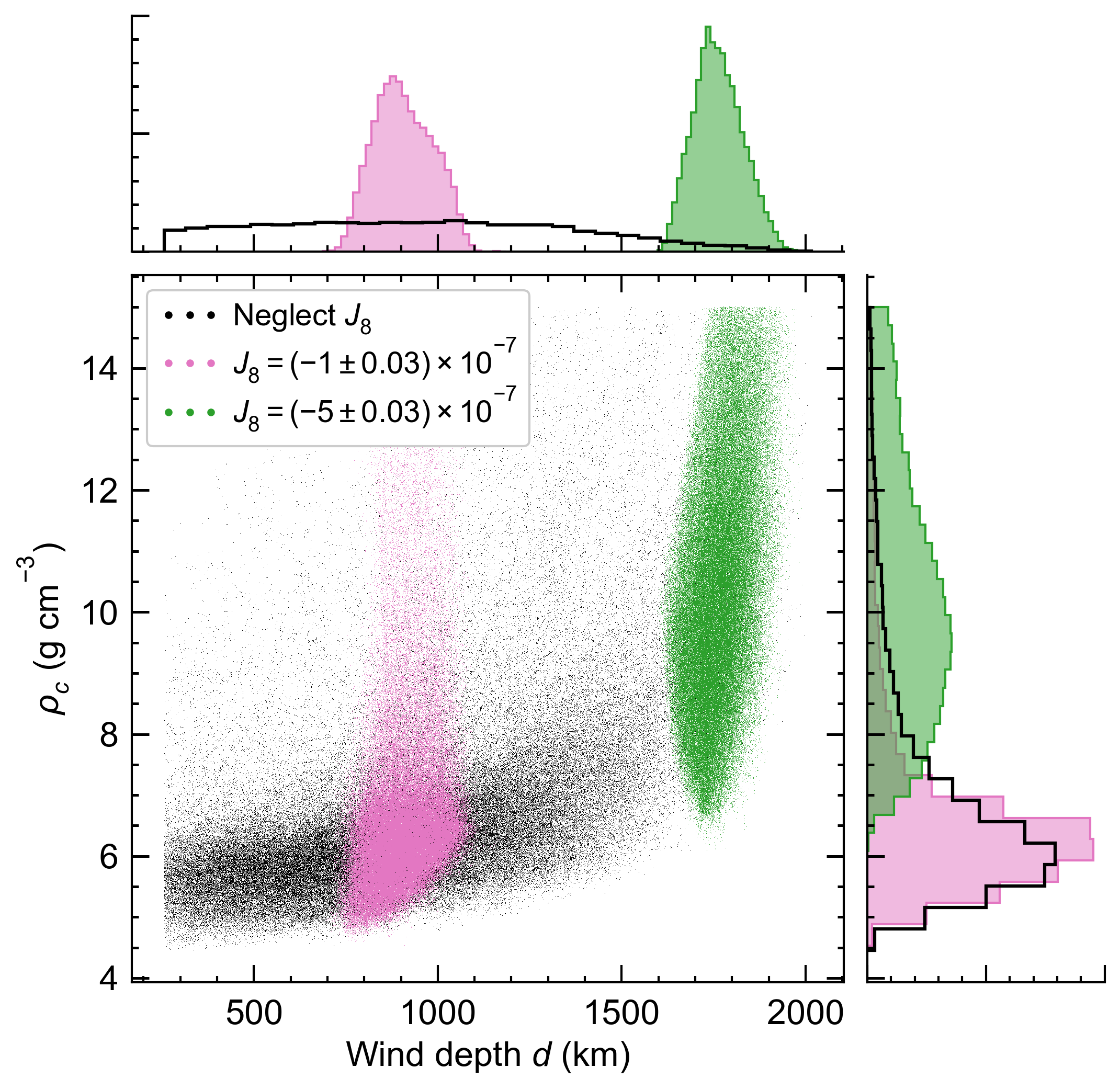}
        \caption{\label{fig.compare_rhoc-depth}
        Comparison of the baseline retrieval to $J_2$ and $J_4$ (black points and histograms) with two retrievals incorporating possible measurements of $J_8$ with values given in the legend (pink and green points/histograms). Results are shown in the space of central density versus wind truncation depth. The $0.03\times10^{-7}$ uncertainties quoted for $J_8$ in the legend reflect the $1\sigma$ standard deviation of the Gaussian likelihood in $J_8$.
        }
    \end{center}
\end{figure}

\subsection{Disentangling winds from internal structure}
\label{sec.results.rock-ice}
Even ignoring the value of studying atmospheric dynamics in its own right, constraining the depth of the winds may be crucial to the premise of gravitational sounding of the planet's interior. From this point of view, the wind depth may be a nuisance parameter, but nonetheless one that is strongly covariant with other interior parameters of interest. Figure~\ref{fig.compare_rhoc-depth} shows each model's central density as a function of wind depth, first for the baseline sample retrieved from $J_2$ and $J_4$ (black points), then for two samples folding in an additional constraint $J_8=-10^{-7}$ (pink points) or $J_8=-5\times10^{-7}$ (green points). Both of the latter samples assign a simple Gaussian likelihood in $J_8$ with $\sigma_{J_8}=3\times10^{-9}$, an uncertainty representative of the two trajectories that pass between Uranus and its main rings (Trajectories 3 and 5 in Figure~\ref{fig.jn_all_fit_j2j4}).

A $J_8$ measurement leads to nearly Gaussian posterior distributions for the wind depth in both cases, yielding $d=(0.035\pm0.009)\,\req=(910\pm230)\,{\rm km}$ assuming the smaller magnitude for $J_8$, and $d=(0.069\pm0.008)\,\req=(1760\pm200)\,{\rm km}$ for the larger magnitude (mean $\pm3\sigma$). Hence, both scenarios lead to a definitive detection of the gravitational signal from Uranus's winds. Meanwhile, considering the central density $\rhoc$ as a simple proxy for deeper interior structure, we find that the scenario implicating deeper winds predicts systematically higher central densities (median density 9.0 g cm$^{-3}$) compared to the scenario implicating shallower winds (median density 5.8 g cm$^{-3}$). The nonzero overlap of the two distributions suggests that a $J_8$ measurement cannot uniquely determine the central density, intuitively unsurprising given this moment's superficially concentrated contribution function \citep{2013Icar..225..548N}. Nonetheless the mostly separate distributions show that, by virtue of eliminating the confounding variable of wind depth, $J_8$ does have discriminating power for the deep interior.

\begin{figure*}[ht!]
    \begin{center}
        \includegraphics[width=0.45\textwidth]{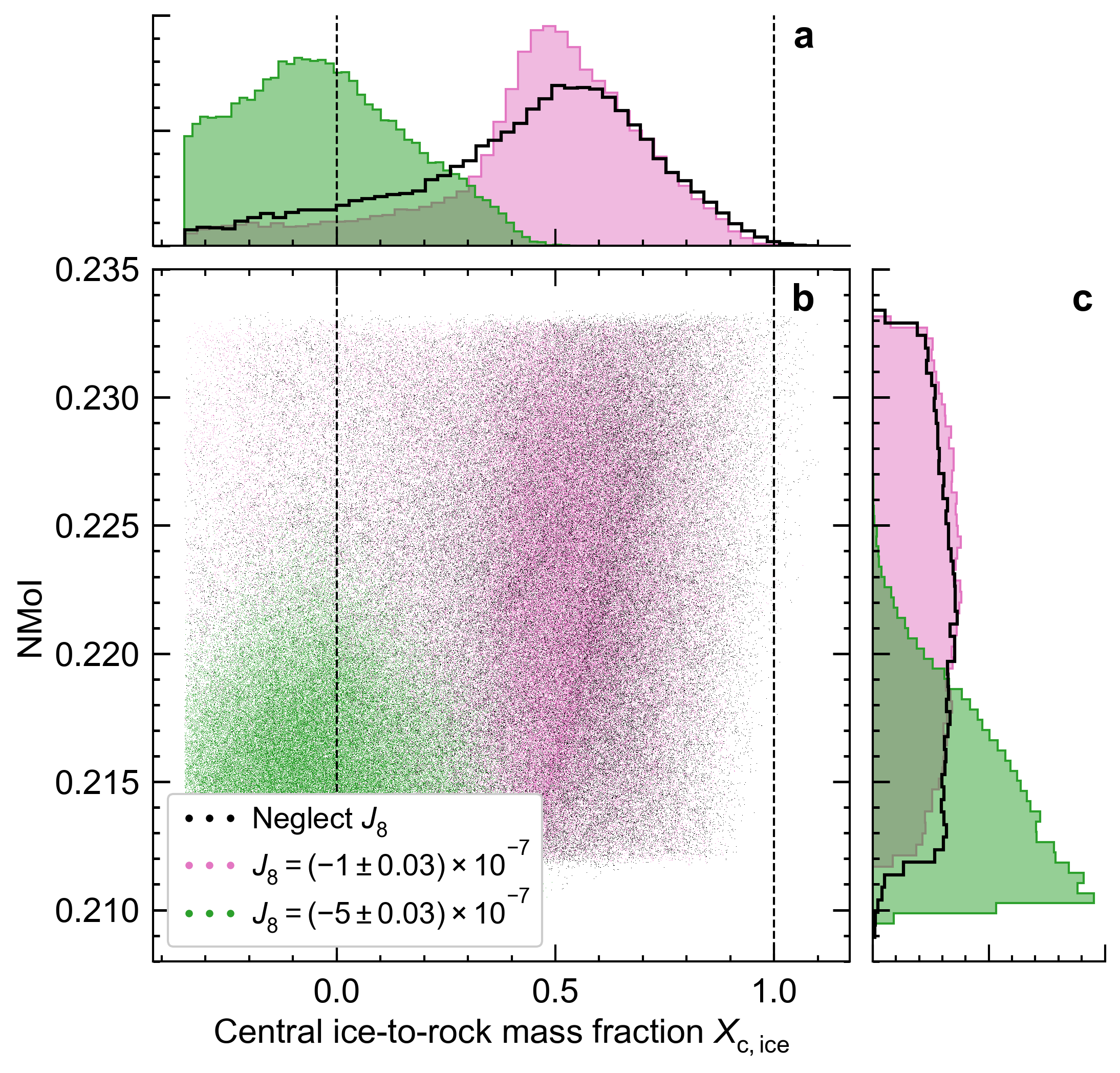}
        \hspace{0.05\textwidth}
        \includegraphics[width=0.45\textwidth]{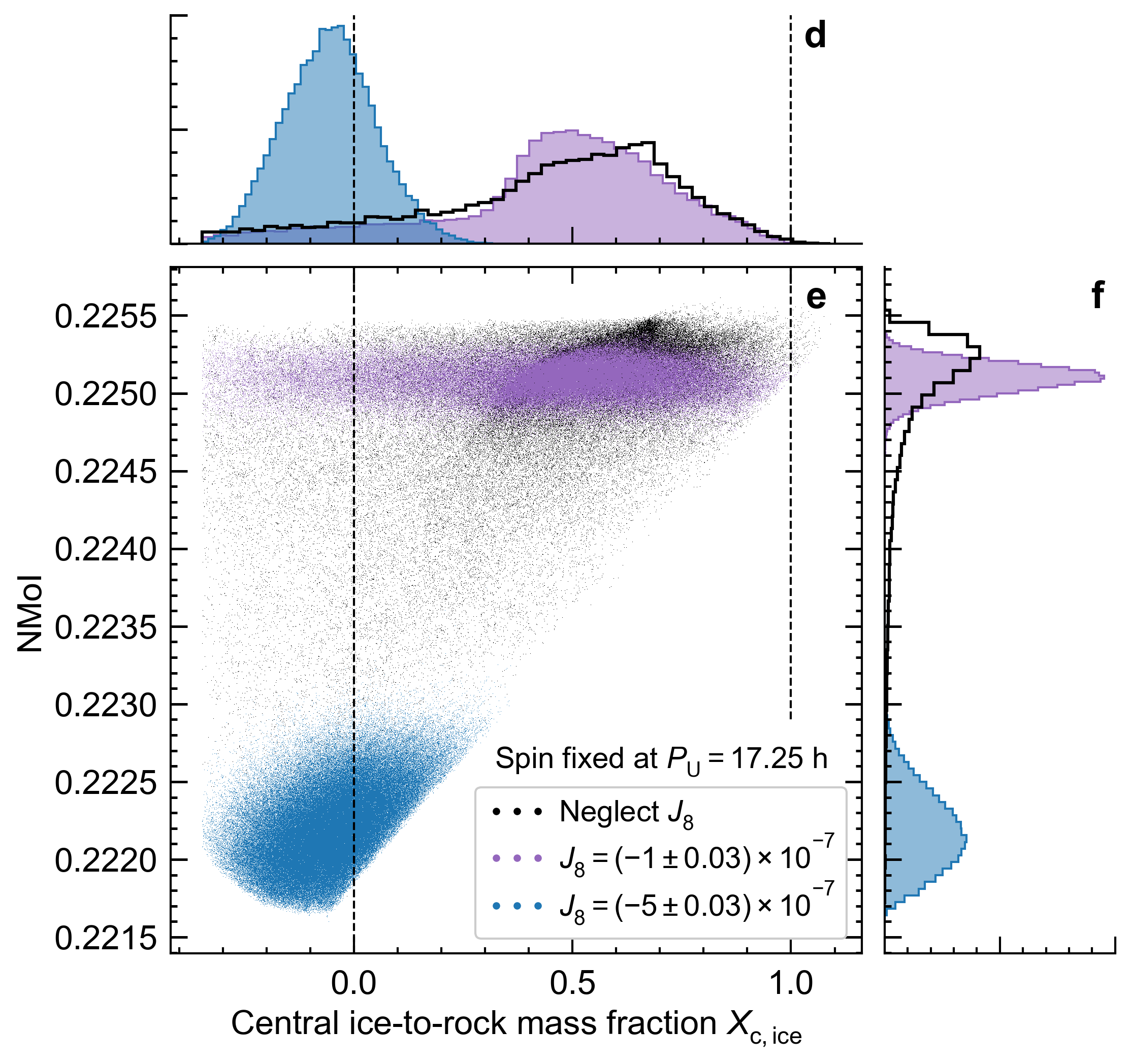}
        \caption{\label{fig.compare_nmoi-xice}
        As in Figure~\ref{fig.compare_rhoc-depth}, but showing the moment of inertia about Uranus's spin axis as a function of the central ice fraction (Equation~\ref{eq.ice_fraction}), from simulated retrievals to different data. Super-unity ice fractions $\xice>1$ correspond to densities lower than that of the adopted ice mixture (see text), suggesting for example dilution by hydrogen. Values close to zero correspond to expectations for pure rock, with values less than zero indicative of a higher fraction of heavier elements like iron. A measurement of $J_8$ leads to a stronger preference for ice- or rock-dominated cores, with the larger magnitude for $J_8$ (green distribution) ruling out ice-dominated compositions entirely. Panels (\textbf{a}-\textbf{c}) show the same samples as in Figure~\ref{fig.compare_rhoc-depth}; panels (\textbf{d}-\textbf{f}) show the corresponding samples that hold the bulk spin period fixed at 17.25~h.
        }
    \end{center}
\end{figure*}

To put the results of Figure~\ref{fig.compare_rhoc-depth} into a more tangible context, we translate the central densities into ice to rock mass ratios post hoc, assuming the \cite{1989Icar...78..102H} pressure-density relations for their representative mixtures of ices (H$_2$O, CH$_4$, NH$_3$) and rocks (SiO$_2$, MgO, FeS, FeO). At the typical central pressures $P_c\approx6.5$~Mbar of our Uranus models, these benchmark densities amount to $\rho_{\rm ice}=4.69$~g~cm$^{-3}$ and $\rho_{\rm rock}=9.57$~g~cm$^{-3}$. Ascribing the density at the model's center to a mixture of these two components alone such that their mass fractions $\xice$ and $X_{\rm rock}$ sum to unity, and working in the additive-volume approximation
\begin{equation}
    \frac{1}{\rhoc}=\frac{X_{\rm rock}}{\rho_{\rm rock}}+\frac{\xice}{\rho_{\rm ice}},
\end{equation}
we infer a central ice mass fraction 
\begin{equation}\label{eq.ice_fraction}
    \xcice = \frac{{\rho_{\rm c,rock}}/\rhoc-1} {{\rho_{\rm c,rock}}/{\rho_{\rm c,ice}} - 1}
\end{equation}
for each model{, with c subscripts denoting the value at the center of the planet}. Since the range of central densities in our models exceeds that of the two end-member component densities, values of $\xice$ outside the range $[0,1]$ are possible. This reflects the reality that Uranus's central regions may include several components with molecular weights outside the limits of the `ice' and `rock' mixtures considered here. In particular, values of $\xice$ exceeding unity would be attainable should the core material be diluted by metallic hydrogen. Values below zero could result from higher concentrations of heavier constituents like iron-bearing compounds. 

{The central density is not necessarily representative of the bulk density in general. The gravity moments' weak sensitivity to the deepest regions of the planet means that a sufficiently small, dense core can generally be added without compromising the fit to the gravity field or strongly affecting the bulk composition. Hence, even an ice-dominated Uranus could have a central density similar to that of rock or iron. However, our gradient model does not allow for this possibility, instead taking the central density to smoothly taper toward hydrogen/helium-like densities in the shallower layers. For this reason, in the context of our models, ice-like central densities ($\xcice\sim1$) do approximate ice-dominated interiors, and likewise $\xcice\sim0$ does approximate a rock-dominated interior. More detailed analyses including temperature dependence and realistic equations of state are warranted, as discussed in Section~\ref{sec.discussion} below.} 

Figure~\ref{fig.compare_nmoi-xice} plots this composition parameter against the nondimensional moment of inertia about the planet's spin axis (\nmoi), calculated as in \cite{2022PSJ.....3...88M}. Panels~a-c show the same three samples as Figure~\ref{fig.compare_rhoc-depth}. Whereas the known gravity field ($J_2$ and $J_4$) yields a nearly uniform distribution in \nmoi, the $J_8$-driven moderate and deep wind scenarios---the pink and green distributions---lead to asymmetric distributions with moderate preference for larger and smaller \nmoi, respectively. Even with the wind depth effectively anchored by the measurement of $J_8$, inference of the moment of inertia is confounded by the uncertain bulk spin of the Uranian interior. Panels~d-f of Figure~\ref{fig.compare_nmoi-xice} show how these distributions are affected if we assume that Uranus's magnetospheric rotation (e.g., \citealt{1986Sci...233...85N,2025NatAs...9..658L}) closely tracks the rotation of the interior; these samples fix the bulk spin period at 17.25~h. The narrow distribution for the \nmoi{} seen in these cases---note the much smaller dynamic range in \nmoi{} in panels~{e} and {f} compared to b and c---reflects the well-defined flattening that would follow from precise knowledge of Uranus's deep spin. 

As for $\xcice$, the distributions returned by the two $J_8$ retrievals are overlapping, yet different enough to show the value of a $J_8$ measurement for testing hypotheses about Uranus's internal composition. Focusing on the marginalized distributions of panel~{a}, we find that the sample with $J_8$ unconstrained leaves rock- and ice-dominated cores approximately equally likely, with 5{3}\% of models having $\xcice$ greater than 0.5. {The scenario with} $J_8=(-1\pm0.03)\times10^{-7}$ (the pink distribution) {again} leads to {approximately half} of models having $\xcice>0.5$. {This follows directly from the similarity of these samples' $\rhoc$ distributions (black and pink histograms in Figure~\ref{fig.compare_rhoc-depth}). However, t}he scenario where deeper winds give rise to a greater magnitude $J_8=(-5\pm0.03)\times10^{-7}$ (the green distribution) {shows a major departure}, favoring a rock-dominated composition ($\xcice<0.5$) at striking {3,000} to 1 odds. Indeed, not a single model is found with $\xcice\gtrsim1$ in this scenario, demonstrating that it is possible on the basis of a $J_8$ measurement to reject the hypothesis that Uranus's deep interior is predominantly water ice by mass. Of course, the detailed conclusions drawn from the data depend on the values of the gravity moments measured and are generally model-dependent, but the rudimentary modeling here is sufficient to motivate the measurement of $J_8$ {as a means of eliminating the influence of zonal winds as a confounding variable}. We note that while fixing the spin period profoundly shrinks the distribution of allowed NMoI, it has little effect on $\xcice$ (Figure~\ref{fig.compare_nmoi-xice}d-e).

\section{Discussion}\label{sec.discussion}
Even at great precision, measurements of $J_2$ and $J_4$ alone are unlikely to elucidate the deep interiors of Uranus or Neptune \citep{2022PSJ.....3...88M}. M25 echoed this finding for Uranus, showing that even optimistic precision gains possible from UOP gravity science yield virtually identical distributions of Uranus's core radius and density. This conclusion holds even if Uranus's differential rotation is negligible, as \cite{2022PSJ.....3...88M} assumed. It is exacerbated, however, if the zonal winds are deep enough to affect the planet's shape and gravity field, leading to major degeneracy and stifling the promise that gravity science holds for the study of structure of Uranus. It is therefore necessary to look to higher degree even, and possibly odd, zonal gravity moments to gain a clearer picture of the planet's interior.

The {odd} zonal moments $J_3$, $J_5$, etc. that probe the mass distribution's equatorial asymmetry are especially attractive because they are a manifestation of the winds alone. These have proven a powerful probe of zonal winds in the gas giants (e.g., \citealt{2018Natur.555..223K,2019GeoRL..46..616G}) and may be similarly useful for Uranus provided that the zonal flow is deep enough for these moments to be detectable. However, the longevity of these north-south asymmetries is unclear{, especially given Uranus's profound seasonal changes as a result of its extreme axial tilt}. Feature tracking shows that latitudes poleward of 60$^\circ$ N exhibit rigid rotation \citep{2015Icar..258..192S}, a stark contrast to the strong shear seen southward of 50$^\circ$ S in Voyager images \citep{2015Icar..250..294K}; see Figure~\ref{fig.baseline_best_windprofs}. \cite{2024Icar..42016186S} studied the hypothesis that this difference reflects a seasonal evolution to the flow near the poles but the evidence is so far inconclusive. Continued observations as Uranus approaches northern solstice{, and high-quality in situ feature tracking by UOP imaging science, will help to} clarify this question.

Should a clear picture of the zonal wind speed profile and its time dependence emerge, $J_3$ through at least $J_9$ would be useful probes of how deep-seated the prevailing flow is, although their magnitudes are limited to levels that would only be measurable by highly inclined orbits with pericenter passages near the equatorial plane and inside the 4, 5, and 6 rings (e.g., T3 in Figure~\ref{fig.baseline_best_jn} or~\ref{fig.jn_all_fit_j2j4}). With measurements this sensitive, we expect that $J_6$ and $J_8$ would be detected as well. Both of these even coefficients are insensitive to the uncertain north-south asymmetries. Among these, $J_6$ is not sensitive to the depth of the winds, but $J_8$ is (Figures~\ref{fig.jn_all_fit_j2j4}~and~\ref{fig.compare_rhoc-depth}).

Our conclusions are based on a highly simplified interior model that does not explicitly account for how composition and temperature enter the density, a suitable approximation for this early stage of planning for the next Uranus mission. Just as for the gas giants, high-pressure equations of state applicable to the interiors of Uranus and Neptune are difficult to constrain experimentally, and much recent work still relies on analytic or semi-analytic equations of state, e.g., ANEOS \citep{aneos} SESAME \citep{sesame}, and QEOS \citep{1988PhFl...31.3059M}; see, e.g., \cite{2013MNRAS.434.3283V,2024A&A...690A.105M,2024A&A...684A.191N}. Descriptions synthesizing ab initio results for H$_2$O are becoming available \citep{2019A&A...621A.128M,2020A&A...643A.105H} and are gaining adoption in planetary modeling (e.g., \citealt{2020A&A...633A..50V,2024ApJ...971..104S,2025ApJ...989L..40T,2025ApJ...988..186A}). The analysis we present here should be revisited with more realistic models,
{particularly to decide whether gravity moments can definitively resolve the debate between a relatively cold and icy interior for Uranus (e.g., \citealt{2024PNAS..12103981M,2025ApJ...990...20M}) or a warm and rocky one \citep{2020A&A...633A..50V,2025ApJ...989L..40T}. Some degree of composition-temperature degeneracy will be inescapable in such an analysis, an aspect sidestepped by the simple temperature-agnostic analysis of Section~\ref{sec.results.rock-ice}. Here we have underscored the fact that the influence of Uranus's winds needs to quantified as a necessary first step, and this can be accomplished by a radio science experiment capable of measuring $J_8$ and possibly the odd $J_n$.}

Another assumption that warrants scrutiny is the idea that Uranus's rotation is uniform below a single shear zone associated with the base of the zonal winds, found here to be limited to pressures $\lesssim10$ kbar based on the observed gravity field. It is thought that the high electrical conductivity of compressed hydrogen keeps the deep interiors of Jupiter and Saturn near perfectly rigid rotation \citep{2008Icar..196..653L,2018Natur.555..227G}. Uranus and Neptune's strongly multipolar magnetic fields have been suggested to arise from dynamo action in convective shells \citep{2004Natur.428..151S}, where the free electrons may be supplied by superionic water \citep{2018NatPh..14..297M,2011Icar..211..798R}. \cite{2024PNAS..12103981M} argue on the basis of ab initio simulations that H$_2$O phase separates from heavier C-, N-, and H-bearing compounds, and the pressure-dependent solubility of H in the latter mixture may naturally yield a stably stratified deep interior (see also \citealt{2025ApJ...990...20M}). Nonetheless all mixtures were found to be good electrical conductors, and hence rigid rotation for the deep interior of Uranus appears likely.

The rotation of Uranus's auroral footprints has been tracked over a long time baseline by \cite{2025NatAs...9..658L}, yielding what appears to be an extraordinarily stable rotation period over decade timescales (see also \citep{1986Natur.322...42D}). The apparent constancy of Uranus's magnetic field rotation stands in contrast to varying magnetospheric rotation observed at Saturn \citep{1981GeoRL...8..253D,2005Sci...307.1255G,2007Sci...316..442G,2006Natur.441...62G}, which suggests imperfect coupling to Saturn's interior rotation \citep{2009Natur.460..608R,2015Natur.520..202H,2019ApJ...879...78M,2019ApJ...871....1M}. We have considered Uranus's deep interior rotation period as a free parameter here, but note that a precise period of 17.25 h for the interior considerably reduces the space of axial moments of inertia permitted by the gravity field (Figure~\ref{fig.compare_nmoi-xice}). A measurement of Uranus's polar 1-bar radius from orbiter radio occultations may aid in confirming this picture for Uranus's rotation state \citep{1987JGR....9214987L,2010Icar..210..446H,2022MNRAS.512.3124N}.

{The feasibility of flying between Uranus's cloud tops and its main rings may hinge on the precise extent and density of the $\zeta$ ring, a faint, extended, dusty ring interior to the 6 ring first imaged by Voyager \citep{1986Sci...233...43S} and subsequently observed from Keck at 2 $\mu$m \citep{2006Icar..180..186D,2010Icar..208..927D,2013Icar..226.1399D}. Based on these more recent observations, \cite{2013Icar..226.1399D} conclude that the $\zeta$ ring's brightness peaks approximately 14,000 km from Uranus's cloud tops, but faint emission also extends far inward, potentially down to the cloud tops themselves. The overall density of dusty material and hazard to a spacecraft traversing close to the planet remain uncertain (see \citealt{2023PSJ.....4..104H}), but images of the ring captured by JWST in 2023 may afford new information \citep{2025PSJ.....6..204H}. Risk to UOP may be mitigated by passing as close to Uranus as is feasible without suffering exessive drag from Uranus's atmosphere (\citealt{2019P&SS..17704680H}; see also discussion in \citealt{2024PSJ.....5..116P}), motivating the 2,500 km periapsis altitude chosen for the trajectories 2 and 3 in Table~\ref{tab.orbits} and Figure~\ref{fig.orbits}.
}

For understanding giant planet atmospheres and interiors, gravity science is most successful when paired with normal mode seismology. Uranus science will benefit from orbits that can sense the multipolar planetary gravity field as well as make observations of Uranus's planetary-scale oscillations in the rings, cloud tops, or dynamical gravity field \citep{2022PSJ.....3..194A,2025PSJ.....6...70M,2020RSPTA.37890475F,friedson25}.

\section{Summary}
We have examined the role that Uranus's zonal flows play in modifying the planet's shape and gravity field. We found that currently available data do not exclude the possibility that these winds are deeply seated enough to confound the interpretation of gravity data to probe Uranus's interior, even if the gravity moments are measured precisely by future missions like the Uranus Orbiter and Probe. 

Assuming that Uranus's deep interior rotates rigidly, separating the influence of this rigidly rotating deep structure from that of the zonal flow demands the measurement of higher-degree even zonal gravity moments ($J_6$ and $J_8$) or odd moments ($J_3$, $J_5$, etc.). {A unique interpretation of the} odd moments {is muddled by} our limited understanding of the persistence of north-south asymmetries in Uranus's zonal flow. {Among these, $J_5$ and $J_7$ appear to be the most robust probes of the depth of Uranus's winds, but their magnitudes may fall below the detection threshold for even the most optimistic orbit (Figure~\ref{fig.jn_all_fit_j2j4}b)}. At degree 6 the gravity field is still dominated by rigid-body oblateness, and hence $J_6$ is a poor probe of the depth of the zonal flow (Figures~\ref{fig.dj2n-depth-spin}-\ref{fig.jn_all_fit_j2j4}). This leaves $J_8$ as the most promising means of deducing the depth of Uranus's winds.

Among the realistic preliminary orbiter trajectories considered here, only eccentric, highly inclined orbits with pericenter passages inside the main rings would be sensitive enough to reveal these gravity moments via Doppler tracking using radio telecommunications. If such an orbit were to fly and measure {odd $J_n$ consistent with zero, and} $J_8$ consistent with {the range expected for rigid rotation}, then deep winds could be ruled out, significantly simplifying the interpretation of these data for Uranus's interior.

\section*{Acknowledgments}
    C.M.'s research was supported by an appointment to the NASA Postdoctoral Program at the Jet Propulsion Laboratory, administered by Oak Ridge Associated Universities under contract with NASA.
    We thank Reza Karimi and Jonathan Lunine for helpful input.
    This work benefitted from ideas and advice from the participants of the September 2023 workshop ``Determining the Interior Structure of Uranus" organized by the W.M. Keck Institute for Space Studies.
    The research described in this paper was carried out at the Jet Propulsion Laboratory, California Institute of Technology, under a contract with the National Aeronautics and Space Administration (80NM0018D0004). 
    JPL internal funding acknowledged. Pre-decisional information, for planning and discussion purposes only.
    \copyright 2025 California Institute of Technology. All rights reserved.


\appendix

\section{Contributions to the odd moments} \label{app.odd_moments}
\begin{figure*}
    \begin{center}
        \includegraphics[width=0.9\textwidth]{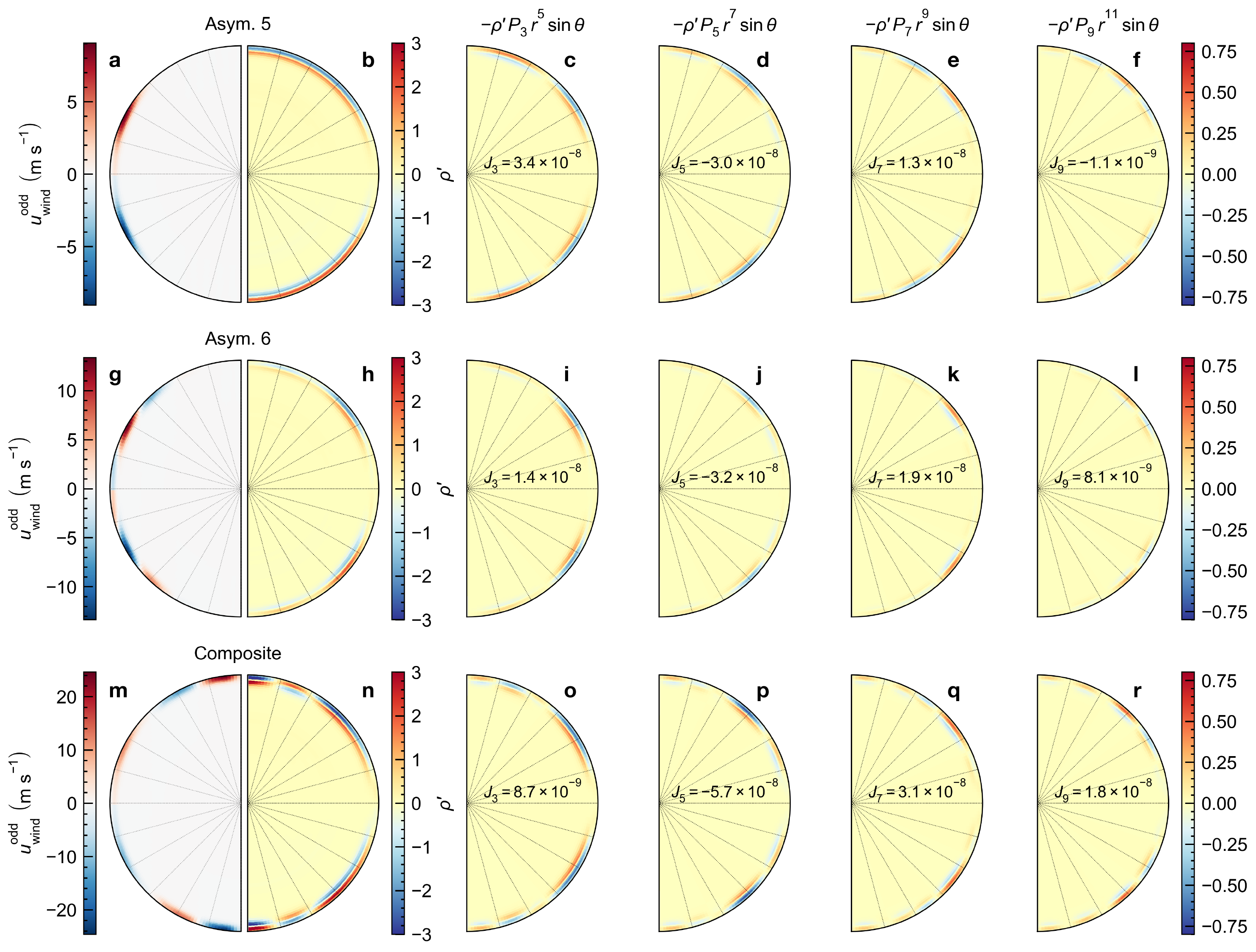}
        \caption{\label{fig.odd_winds_jn_integrands}
        Contributions to the odd gravity moments in the interior model from Figure~\ref{fig.baseline_best_jn}. Each row assumes a different tropospheric wind profile.
        The first row assumes the profile \textit{Asym. 5}, showing meridional slices of the odd part of the asymmetric wind profile (panel \textbf{a}) and the induced perturbation to mass density (panel \textbf{b}). Panels \textbf{c}-\textbf{f} show the integrands in Equation~\ref{eq.jn} for $J_3$, $J_5$, $J_7$, and $J_9$ respectively, including the factor of $r^2\sin\theta$ from the spherical volume element, as well as an overall minus sign so positive values in the colormap contribute to $J_n$ in the positive sense. Panels \textbf{g}-\textbf{l} assume the profile \text{Asym. 6} and panels \textbf{m}-\textbf{r} assume the Composite profile. Faint spokes mark $15^\circ$ increments. 
        All density perturbations (blue-yellow-red colormaps) are in units of $10^{-5}\ {\rm g\ cm}^{-3}$ and $r$ is in units 25,559 km.
        }
    \end{center}
\end{figure*}
Section~\ref{sec.results} shows that the odd $J_n$, especially $J_3$ and $J_9$, are notably sensitive to the assumed wind profile (see Figures~\ref{fig.baseline_best_jn},~\ref{fig.jn_all_fit_j2j4}b, and~\ref{fig.j3}). Since the disparity in these odd $J_n$ limits their utility for measuring the depth of Uranus's winds, we examine the contributions that the assumed flow profile makes to the gravity moments as a function of latitude. 

Figure~\ref{fig.odd_winds_jn_integrands} presents intermediate results in the calculation of the odd $J_n$ for the same interior model highlighted in Figure~\ref{fig.baseline_best_jn}. In addition to the odd part of the wind speeds and the wind-induced density perturbation, the figure shows this dynamical density perturbation $\rho^\prime(r,\mu)$ weighted by the factor of $r^n P_n(\cos\theta)$ that appears in the expression for the zonal gravity moments:
\begin{equation}\label{eq.jn}
    J_n=-\frac{1}{MR_{\rm eq}^n}\int\rho^\prime(r,\theta)r^{n}P_n(\cos\theta)\,r^2\sin\theta\,dr\,d\theta\,d\phi\quad(n\ {\rm odd}).
\end{equation}
These panels include the factor of $r^2\sin\theta$ from the spherical volume element to give a direct visual indication of these moments' sensitivity to the zonal flow, analogous to the one-dimensional `contribution functions' invoked to describe the static $J_n$ arising from oblateness \citep{2013Icar..225..548N}.

The long latitudinal wavelength of the $n=3$ Legendre function invites more self-cancellation within each hemisphere for the \textit{Asym. 6} and \textit{Composite} profiles, which have finer latitudinal structure. Hence, the most slowly varying and most equatorially concentrated profile \textit{Asym. 5} yields the largest magnitudes for $J_3$, as is the overwhelming trend in our retrievals seen in Figures~\ref{fig.jn_all_fit_j2j4}-\ref{fig.j3}. The spatial coherence of this profile within each hemisphere leads the odd moments to decline in magnitude toward higher degree due to self-cancellation. In the \textit{Asym. 6} and \textit{Composite} cases, the variety of latitudinal scales present in the wind profile, and the extent of the velocity to higher latitudes, produce a nontrivial relationship between the moment magnitude and degree $n$, with $|J_n|$ peaking at $n=5$ in the \textit{Asym. 6} case and at $n=7$ for the \textit{Composite} case. Of the odd moments considered here, $J_9$ shows the strongest dependence on the wind profile, apparently mostly driven by the latitudinal shear between 15 and 60$^\circ$ latitude, which happens to be effectively sampled by one full cycle of $P_9(\mu)$. The final value of $J_n$ is then dictated by the details of the zonal flow in this region. 

The broadly distributed nature of the Legendre functions means that the conventional use of the expansion (Equation~\ref{eq.jn}) can make it difficult to ascribe a single part of the atmospheric flow (e.g., a single latitudinal band) to each moment $J_n$, although numerical experiments isolating regions of the flow (see for instance \citealt{2025NatCo..16.2618G}) can be instructive. Resolving the gravitational influence of more localized features may be aided by analyzing Doppler data using Slepian functions instead \citep{2025GeoRL..5213236K,2020JGRE..12506416P,2019ApJ...874L..24G}.

Clearly the details of the odd $J_n$ spectrum are highly sensitive to the assumed wind profile, motivating increased attention to tracking Uranus's tropospheric features from the ground, and ideally culminating in orbiter-based imaging covering a broad range in latitude.

\end{document}